\documentclass[preprint]{aastex}
\input psfig.sty

\shortauthors{J.P. Maier, G.A.H. Walker, D.A. Bohlender}
\shorttitle{Interstellar C$_{4}$ and C$_{5}$}

\begin{document}

\title{LIMITS TO INTERSTELLAR C$_{4}$ AND C$_{5}$ TOWARDS
$\zeta$~OPHIUCHI}

\author {John P. Maier\altaffilmark{1} }
\affil{Institute for Physical Chemistry, Klingelbergstrasse 80,
University of Basel CH-4053, Switzerland} \email{j.p.maier@unibas.ch}

\altaffiltext{1}{Visiting Astronomer, Canada-France-Hawaii Telescope,
operated by the National Research Council of Canada, the Centre
National de la Recherche Scientifique of France, and the University of
Hawaii.}

\author { Gordon A.H. Walker\altaffilmark{1}} \affil{1234 Hewlett Place, Victoria, BC, Canada V8S 4P7} \email{walker@astro.ubc.ca}

\author {David A. Bohlender\altaffilmark{1}} \affil{National Research
Council of Canada, Herzberg Institute of Astrophysics\\ 5071 West
Saanich Road Victoria, BC, Canada V9E 2E7}
\email{david.bohlender@nrc.ca}

\begin{abstract} 
We have made a sensitive search for the origin bands in the known
electronic transitions of the linear carbon chains C$_{4}$  and C$_{5}$
at 3789 and 5109 \AA\/ towards $\zeta$ Oph (A$_{V}$$\leq$1). The
incentive was a recent detection of C$_{3}$ in this interstellar cloud
with a column density of 1.6$\times$10$^{12}$ cm$^{-2}$ plus the
availability of laboratory gas phase spectra of C$_{4}$  and C$_{5}$.
Further, some models of diffuse interstellar clouds predict that the
abundance of these latter species should be within an order of
magnitude of C$_{3}$. Despite achieving S/N of 2300 to 2600 per pixel
at a resolution of $\sim$110,000, the searches were negative, leading
to 3$\sigma$  upper limits to the column density of N(C$_{5}$)=
2$\times$10$^{11}$cm$^{-2}$ and N(C$_{4}$)= 4$\times$10$^{12-13}$
cm$^{-2}$ where these values rely on theoretically calculated
oscillator strengths. The implication of these limits are discussed on
the choice of molecules for study in future attempts to identify the
carriers of the stronger diffuse interstellar bands.

\end{abstract}


\section{Introduction} Carbon chain molecules are often at the
forefront in discussions of the diffuse interstellar bands (DIB) (see
\citet{Dou77} and \citet{Smi77}), which are found mainly in the optical
part of the spectrum (Herbig 1995). They became more appealing
candidates with the discovery at mm wavelengths \citep{Mcc01} of
numerous polar carbon chains in dense interstellar clouds. However,
only in the past few years have the gas phase electronic spectra of a
number of such species  (e.g.  C$_{4}$, C$_{5}$, C$_{2n}$H n=3-6), as
well of related cations (e.g.  HC$_{2n}$H$^{+}$  n=2-4,
HC$_{2n}$N$^{+}$  n=2-4) and anions (e.g.  C$_{n}^{-}$  n=3-11) been
detected in the laboratory, enabling direct comparison with
astronomical data (for example, \citet{Mot99}). In all cases where the
comparisons could be made, the results were negative, leading to the
conclusion that the column densities of these species are $\leq
10^{12}$ cm$^{-2}$ in diffuse clouds.

However, with the detection of the rotational lines in the electronic
transition of C$_{3}$  near 4052 \AA\/  \citep{Mai01}, corresponding to
total column densities of 1-2$\times$10$^{12}$ cm$^{-2}$,  and the
prediction of certain models of such diffuse clouds (e.g.
\citet{Ter98}) that the abundances of C$_{4}$ and C$_{5}$ are less than
a factor of ten smaller than that of C$_{3}$, it became appealing to
search for these molecules. A disadvantage compared to the electronic
spectrum of C$_{3}$ is that, according to theoretical predictions, the
oscillator strengths of the transitions appear to be smaller.
Unfortunately, experimentally determined oscillator strengths are
unavailable. These smaller oscillator strengths result in higher
abundance limits than was the case for C$_{3}$. Apart from the
abundances predicted for C$_{4}$ and C$_{5}$,  the detection of both
C$_{3}$ and C$_{5}$ in the infrared by \cite{Ber89} in a circumstellar
shell provides a guideline. In the latter work the concentration of
C$_{5}$ proved to be some ten times less than that of C$_{3}$.

Here we report an attempt to detect the origin bands of
the electronic transitions of C$_{4}$ ($^{3}\Sigma_{u}^{-}$ $-$
$^{3}\Sigma_{g}^{-}$ ) at 3789 \AA\/ and of C$_{5}$  ($^{1}\Pi_{u}$ $-$
$^{1}\Sigma_{g}^{+}$ ) at 5109 \AA\/ in absorption towards the
reddened star $\zeta$ Oph (HD 149757) with the Gecko spectrograph of the
Canada-France-Hawaii 3.6 m telescope. There is already one report in the
literature of a non-detection of C$_{5}$  \citep{Gal01}. The
present study is an order of magnitude more sensitive. There is no
prior attempt reported  to detect C$_{4}$ in
interstellar clouds, but the electronic spectrum in the gas phase was
only obtained a year ago \citep{Lin00}.

\section{The Observations}

The reddened star $\zeta$ Oph (HD 149757), was observed on 29 and 30
June 2001 (UT) with the Gecko echellette spectrograph, fiber fed from
the Cassegrain focus of the Canada-France-Hawaii 3.6-m telescope (CFHT)
\citep{Bau00}. This star is bright, having a visual extinction, A$_{v}$, near
1 with a rich spectrum of sharp interstellar lines. \cite{Cra97} has
resolved the interstellar C$_{2}$ at 8756 \AA\/ into two close velocity
components separated by 1.1 km s$^{-1}$ which is hard to resolve at
our resolution. The rapidly rotating star, $\eta$ UMa (HD 120315) was
observed as standard.

The detector was the rear illuminated EEV1 CCD (13.5 $\mu$m$^{2}$
pixels) and the spectral regions were centered at 3789 \AA\/ in the
15th order, and at 5109 \AA\/ in the 11th order. The 15th order was
isolated by the Gecko ultraviolet grism, the 11th by the blue grism.
Individual spectra of $\zeta$ Oph were exposed for 15 minutes at 3789
\AA\/ and 6 minutes at 5109 \AA\/. The feed fibre was continuously
agitated to overcome modal noise (see \cite{Bau01}).  Lines in 
Th/Ar comparison arc spectra, taken before and after each set of
stellar spectra, typically had FWHM of 3 pixels, corresponding to
resolutions of $R$ = 115000 and 109800 at 3789 and 5109 \AA, or 0.0109
and 0.0155 \AA\/ pxl$^{-1}$, respectively. Extensive series of
flat-field spectra of a quartz-iodide lamp were recorded for each
spectrograph setting and groups of biases were taken several times each
night.

Conditions were not ideal to achieve the very high S/N ($>$4000) which
we felt necessary to detect the C$_{4}$ and C$_{5}$ bands. Only parts
of two of the assigned nights were clear and the flat-field spectra
displayed an unstable structure typical of modal noise. The stellar
spectra did not show the same structure and we were able to demonstrate
that the main fiber agitation was working as designed but a
misalignment of the fiber feeding the the light of the flat field lamp
to the main fiber may have caused the problem. For this reason a
combination of flat fields and standard stellar spectra were used in
the reduction.

The many biases taken throughout the observing run were averaged to remove
the zero-level offset in the spectra for both wavelength regions.  For the
5109 \AA\/ spectra, obtained after the noise problem with the flat
field exposures had been recognized, $\eta$ UMa was observed at a S/N level
substantially higher than $\zeta$ Oph.  Since the spectrum of the standard star
is featureless near 5109 \AA\/ its spectrum was used in place of flat
field spectra.

Unfortunately, at 3789 \AA, weather constraints prevented observations of $\eta$ UMa at a sufficient S/N level for them to be used directly as flat fields.  Instead, we averaged the flat field exposure to remove pixel-to-pixel
sensitivity variations in the spectra of both $\eta$ UMa and $\zeta$ Oph.
One dimensional spectra of both stars were then extracted in a standard
manner.  Because of the modal noise in the flat field spectra, each flat field corrected stellar spectrum shows a residual low frequency modal
noise pattern.  We therefore smoothed the featureless, flat fielded spectrum of
$\eta$ UMa with a 10-point box car filter and then divided the $\zeta$ Oph
spectrum by this smoothed spectrum to remove the modal noise residuals without
seriously degrading the inherently high S/N of the $\zeta$ Oph spectrum.  For
clarity this reduction procedure is illustrated in Figure 1.

Low-order polynomial fits to the positions of the Th/Ar arc spectra were used to calibrate the $\zeta$ Oph spectra in
wavelength.  The spectra were then normalized to the continuum and
heliocentric corrections applied. The observations are summarised in Table \ref{observations} which lists
exposure times and S/N per pixel for each spectral region. The final
column gives the radial velocities of the interstellar K I 4044.1 and
4047.2 \AA\/ lines as quoted by \cite{Mai01}. These velocities were
applied to each spectrum to put the interstellar features on a
laboratory scale before making the comparisons discussed in the next
section. The comparisons are shown in Figures 2 and 3 where the stellar spectra
have been smoothed with a 3-pixel box car filter.

The 3$\sigma$ detection
limits are derived from:

\[W_{max} = 3(w d)^{\frac{1}{2}} (S/N)^{-1}\]

\noindent where the 3$\sigma$ limiting equivalent width, $W_{max}$, and
the FWHM of the feature, $w$, are both measured in \AA, the
spectrograph dispersion, $d$, in \AA\/ pixel$^{-1}$, and S/N is the
signal to noise per pixel. We adopted $w$ = 0.24 and 0.13 for the 3789
and 5109 \AA\/ laboratory features, respectively.

\begin{deluxetable}{ccccccccc}
\tabletypesize{\footnotesize}
\tablecaption{The Observations of $\zeta$ Oph (HD 149757) \label{observations} }
\tablewidth{0pt}
\tablehead{  
\multicolumn{3}{c}{3789 \AA} &&
\multicolumn{3}{c}{5109 \AA} &&
\colhead{K I Rad.Vel\tablenotemark{d}}\\ \cline{1-3} \cline{5-7}
\colhead{T\tablenotemark{a}} & \colhead{S/N\tablenotemark{b}} & 
\colhead{$W(10^{-4}$\AA)\tablenotemark{c}} &&
\colhead{T\tablenotemark{a}} & \colhead{S/N\tablenotemark{b}} &
\colhead{$W(10^{-4}$\AA)\tablenotemark{c}} &&
\colhead{(km s$^{-1}$)}
}
\startdata
 11700 & 2300 &0.67& &10800 & 2600 &0.50&& $-$14.53 $\pm$0.18 \\ 
 \enddata
\tablenotetext{a}{Total exposure times in seconds.}
\tablenotetext{b}{per pixel}
\tablenotetext{c}{3$\sigma$ equivalent width detection limit for the band head (see section 2)}
\tablenotetext{d}{from \cite{Mai01}}
\end{deluxetable}

\section{ Results and Discussion}

Our attempt to detect the C$_{4}$  3789 \AA\/ origin band of the known electronic systems in
absorption through the diffuse cloud towards the reddened star
$\zeta$ Oph was negative. In Figure 2 we compare the  stellar spectrum
with one from the laboratory \citep{Lin00}  recorded at a temperature
of around 50 K. This temperature should be representative for a
non-polar molecule in diffuse interstellar clouds; in the case of C$_{2}$
\citep{Lam95} and C$_{3}$ \citep{Mai01}, comparable temperatures have
been inferred. The laboratory spectrum (which differs from the
published one only in that the overlapping C$_{2}$ lines have been
removed) shows that the rotational lines are lifetime broadened
allowing only the lines in the P-branch to be resolved. The intense
part of the band to the blue is the R-head and this feature was
primarily sought in the astronomical spectrum. The 3$\sigma$ upper limit
for the equivalent width ($W_{max}$) detection of the R-head is
0.67$\times$10$^{-4}$ \AA. The upper limit to the column density $N_{max}$
can then be derived from:

\[N_{max} = 1.13\times 10^{20} W_{max}/\lambda^{2}f ~~~{\rm cm}^{-2}\] 

\noindent where $f$ is the oscillator strength of the band at
wavelength $\lambda$ in \AA. 

Though the $^{3}\Sigma_{u}^{-}$ $-$ $^{3}\Sigma_{g}^{-}$ electronic
transition of C$_{4}$ has been observed in the gas phase, the
oscillator strength $f$  is not known experimentally. Two values from
ab initio calculations are available which, however, differ by an order
of magnitude. The first study yielded $f$=0.003 \citep{Pac88} whereas the second gave
f=0.0005 \citep{Mul00}. The values are for the whole band
system and thus the $f_{0-0}$ for the 3789 \AA\/ band will be reduced
by the Franck-Condon factor for this transition. Based on the intensity
distribution of the vibrational bands observed in the absorption
spectrum measured in a neon matrix \citep{Fre96}, this may be a factor of
5. Thus, taking $f_{0-0}$ in the 0.0001 to 0.001 range leads to N(C$_{4}$)
$\leq4\times$ 10$^{12}$ to 10$^{13}$ cm$^{-2}$.
   
Two more recent models of diffuse interstellar clouds (of which the
data have been made available to the authors) with characteristics
comparable to the one towards $\zeta$ Oph \citep{vanD86} yield
N(C$_{3}$)/ N(C$_{4}$) ratios of about 2 \citep{Ter98}, and 13
\citep{Tur00}. The detected total column density of C$_{3}$ is
1.6$\times$10$^{12}$ cm$^{-2}$ \citep{Mai01}. Thus, in view of the
uncertainty in the oscillator strength, it is not possible to assess
the predictive value of the models.  

The  situation is similar for C$_{5}$. The band system with the 5109
\AA\/ origin band shown in Figure 3 was observed first in absorption in a
5 K neon matrix \citep{For96} and then in the gas phase \citep{Mot99}.
It was assigned as the $^{1}\Pi_{u}$ $-$ $^{1}\Sigma_{g}^{+}$
electronic transition by analogy to the comet band system of C$_{3}$.
The rotational structure is only partially resolved due to line widths
of  0.7 cm$^{-1}$ (homogeneous broadening due to intramolecular
processes). The  oscillator strength $f_{0-0}$ = 0.02  was estimated by
taking the experimentally known value for C$_{3}$ ($f_{0-0}$ = 0.016)
and scaling this up by the length of the molecule, as simple quantum
models such as particle in a box predict.

However, on the basis of  recent theoretical calculations, it is
proposed that the 5109 \AA\/ band system is in fact a forbidden
transition, to $^{1}\Delta_{u}$ and/or  $^{1}\Sigma_{u}^{-}$  states
\citep{Han01}. The oscillator strength could not be calculated because
the vibronic effects which lead to its intensity would have to be
correctly accounted for.

There is another band system in the neon absorption spectrum with
origin band at 4454 \AA\/ (unpublished data from the laboratory at
Basel), which would then be the $^{1}\Pi_{u}$ - $^{1}\Sigma_{g}^{+}$
electronic transition, with calculated oscillator strength of 0.03
\citep{Han01}; an earlier f value for this transition was 0.037
\citep{Kol95}. A pragmatic approach to the estimation of the oscillator
strength for the 5109 \AA\/ band is to use the relative intensities of
the absorption systems in the 5 K neon matrix. The 4454 \AA\/
absorption band system appears about a factor of 5 more intense than
the 5109 \AA\/ one, implying an f value for the system of around
0.006.  The $f_{0-0}$  value for the 5109 \AA\/ band will then be
reduced further by its Franck-Condon factor, to yield $f_{0-0}$  around
0.001.  The lower spectrum in Figure 3 is that recorded towards $\zeta$
Oph.  The signal-to-noise ratio (2600) is high, leading to a 3$\sigma$
detection limit of 5$\times$10$^{-5}$ \AA. As the laboratory spectrum
corresponds to a temperature of around 50 K, about 10 rotational lines
comprise the Q-band head.  This means that our astronomical
measurements had a detection limit for an individual rotational line of
$\sim5\times10^{-6}$ \AA. For a lower temperature of say, 10 K, there
would still be some five lines unresolved within the band to give a
3$\sigma$ limit per line of $\sim10^{-5}$ \AA.

Our 3$\sigma$ detetection limit of $\sim5\times10^{-5}$ \AA\/ for the
band, together with $f_{0-0}$ = 0.001, leads to N(C$_{5}$)
$\leq$2$\times$10$^{11}$ cm$^{-2}$.  \cite{Gal01} recently reported a
value of N(C$_{5}$) $\leq10^{11}$ cm$^{-2}$, however the authors were
unaware of the spectroscopic problems associated with the oscillator
strengths and used a value of $f_{0-0}$ of 0.02. Taking the presently
suggested $f_{0-0}$ =0.001 increases their value to N(C$_{5}$) $\leq$
2$\times$10$^{12}$ cm$^{-2}$, which is an order of magnitude less
sensitive than our results.

\subsection{Conclusion}

According to the results reported here, N(C$_{4}$) $\leq$
4$\times$10$^{12}$ to 10$^{13}$ cm$^{-2}$, and N(C$_{5}$) $\leq$
2$\times$10$^{11}$ cm$^{-2}$ in the diffuse cloud towards $\zeta$ Oph,
while N(C$_{3}$) = 1.6$\times$10$^{12}$ cm$^{-2}$ from the earlier
study \citep{Mai01}.  From these column densities we calculate
abundances relative to the total column density of hydrogen (n(H) +
2n(H$_{2}$)), 1.3$\times10^{21}$ cm$^{-2}$ (from Table 2 of
\cite{vanD86}), of C$_{3}$ = 2$\times10^{-9}$, C$_{4}$
$\leq$4$\times10^{-9}$  to 10$^{-10}$, and C$_{5}$ $\leq$
2$\times$10$^{-10}$. The results of the two diffuse cloud models to
which we have access yield abundances which are too high for C$_{3}$ by
about an order of magnitude, 3$\times$10$^{-8}$  \citep{Ter98}, and
5$\times$10$^{-8}$ \citep{Tur00}. The values from the former model
correspond to quasi steady state after 10$^{5}$ years and the latter
are for A$_{V}$ = 1 and n= 500 cm$^{-1}$ chosen with depletions which
provide a good fit for essentially all molecular species in translucent
clouds and for the majority of species in diffuse clouds. Similarly,
the predicted abundances are also too high for the longer chains with
relative values of C$_{3}$:C$_{4}$:C$_{5}$ of 5:3:8 from \cite{Ter98},
although \cite{Tur00} gives 25:2:1.  So it appears that
photodissociation rates have been underestimated (or other depletion
mechanisms such as  electron attachment or  absorption on grains need
to be included). The upper limit we have obtained of 0.1 to the
C$_{5}$:C$_{3}$ ratio is consistent with the result of infrared
detection of both these species in a circumstellar shell with a ratio
of 0.09 \citep{Ber89}, suggesting that our measurements may have been
close to actually detecting C$_{5}$.

The detection of polyatomic species containing carbon atoms in diffuse
and translucent interstellar clouds indicate comparable column
densities, not larger than about 10$^{12}$ to 10$^{13}$ cm$^{-2}$. This
set comprises the detection of C$_{3}$ in the optical region
\citep{Mai01} as well as of the rotational spectra of the polar
species, C$_{2}$H,  C$_{3}$H$_{2}$ \citep{Luc00} as examples. So far
among polyatomic species only H$_{3}^{+}$ has column densities in the
diffuse medium exceeding 10$^{14}$  cm$^{-2}$   \citep{McC98}. The
implication of this for the search of appropriate molecular systems
which could correspond to the stronger, narrower DIBs, with typical
equivalent widths of 0.1 A, is as follows. Assuming that the species
would have oscillator strengths of electronic transitions (in the
visible) in the 1-10 range, as could be the case for carbon chains with
10-20 atoms, then the column density would have to be about 10$^{11}$
to 10$^{12}$  cm$^{-2}$. This column density could be easily attained
by the larger carbon species, whatever their shape, because they  are
less efficiently photodissociated.

The suggestion by \cite{Dou77} that long carbon chain molecules,
C$_{n}$ ($n$ = 5-15) be considered as the carriers of DIBs, has now
been tested directly for C$_{4}$ and C$_{5}$, showing that their
abundance is too small. Thus the next step in attempting to identify
carriers of the DIBs would be to obtain laboratory spectra of carbon
species with 10-20 atoms having electronic transitions in the visible
region (e.g., as is the case for C$_{2n+1}$ n=8-20, \cite{Mai98}) and
then making a direct comparison with astronomical observations at high
S/N as has been demonstrated in this work.

\acknowledgements

Support of the Swiss National Science Foundation (project 20-63459.00)
(J.P.M.), the Canadian Natural Sciences and Engineering Research
Council (G.A.H.W.) and the National Research Council of Canada (D.A.B.)
is gratefully acknowledged.

\clearpage

\begin{figure}
\centerline{\psfig{figure=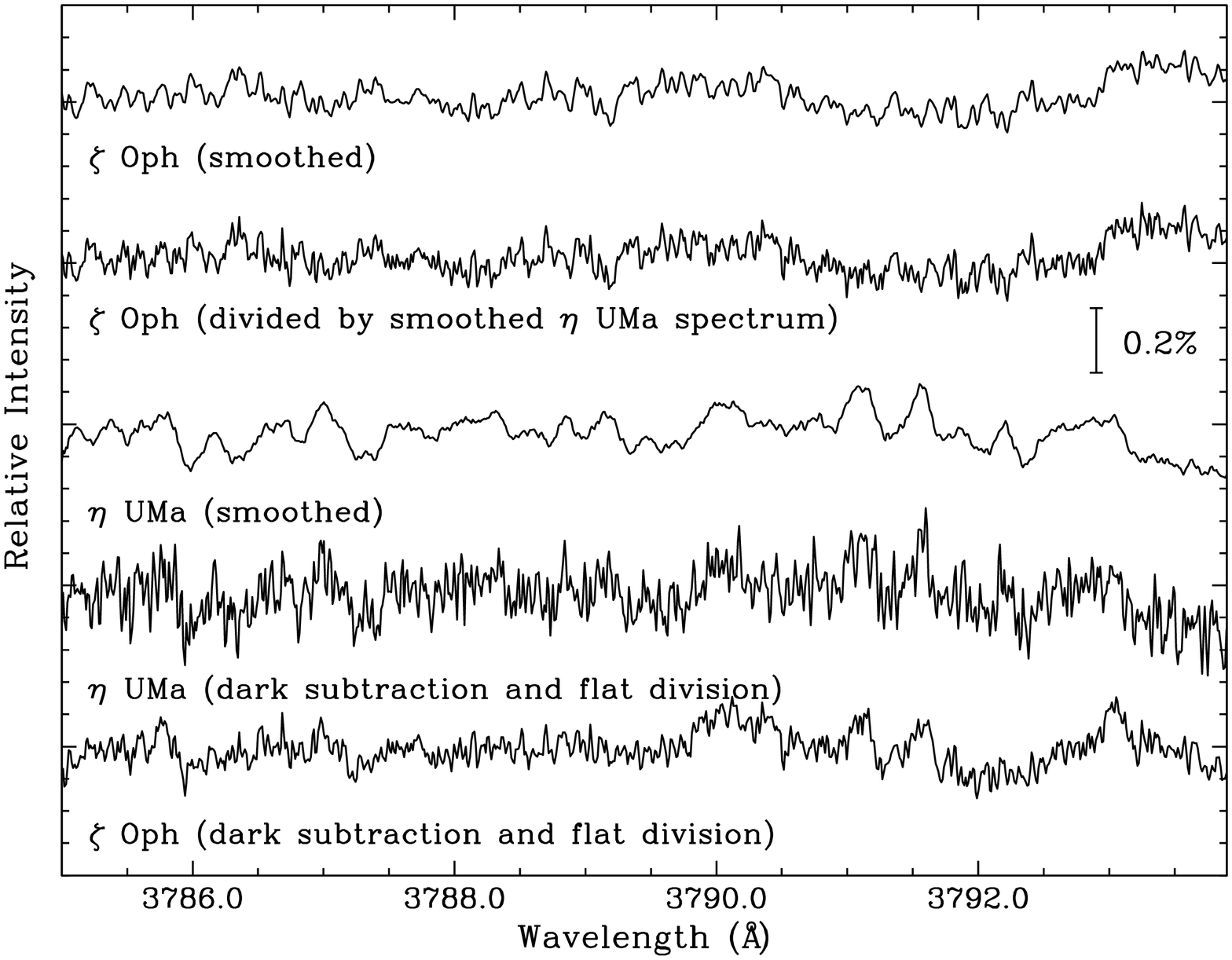,width=18cm,angle=-90}}
\caption{
The final reduction sequence for the spectra at 3790 \AA. The flat
field spectra contained a coarse structure absent from the stellar spectra
which can be seen as common residuals in the dark subtracted and
flat-fielded spectra for $\zeta$ Oph and $\eta$ UMa in the two bottom
plots. The processed spectra of $\eta$ UMa were smoothed and then
divided into the processed spectra of $\zeta$ Oph thereby largely removing the
flat-field structure. The final 3-point smoothed spectrum of $\zeta$
Oph, the one discussed in the paper, is shown in the top plot.
\label{figure1} }
\end{figure}

\begin{figure}
\centerline{\psfig{figure=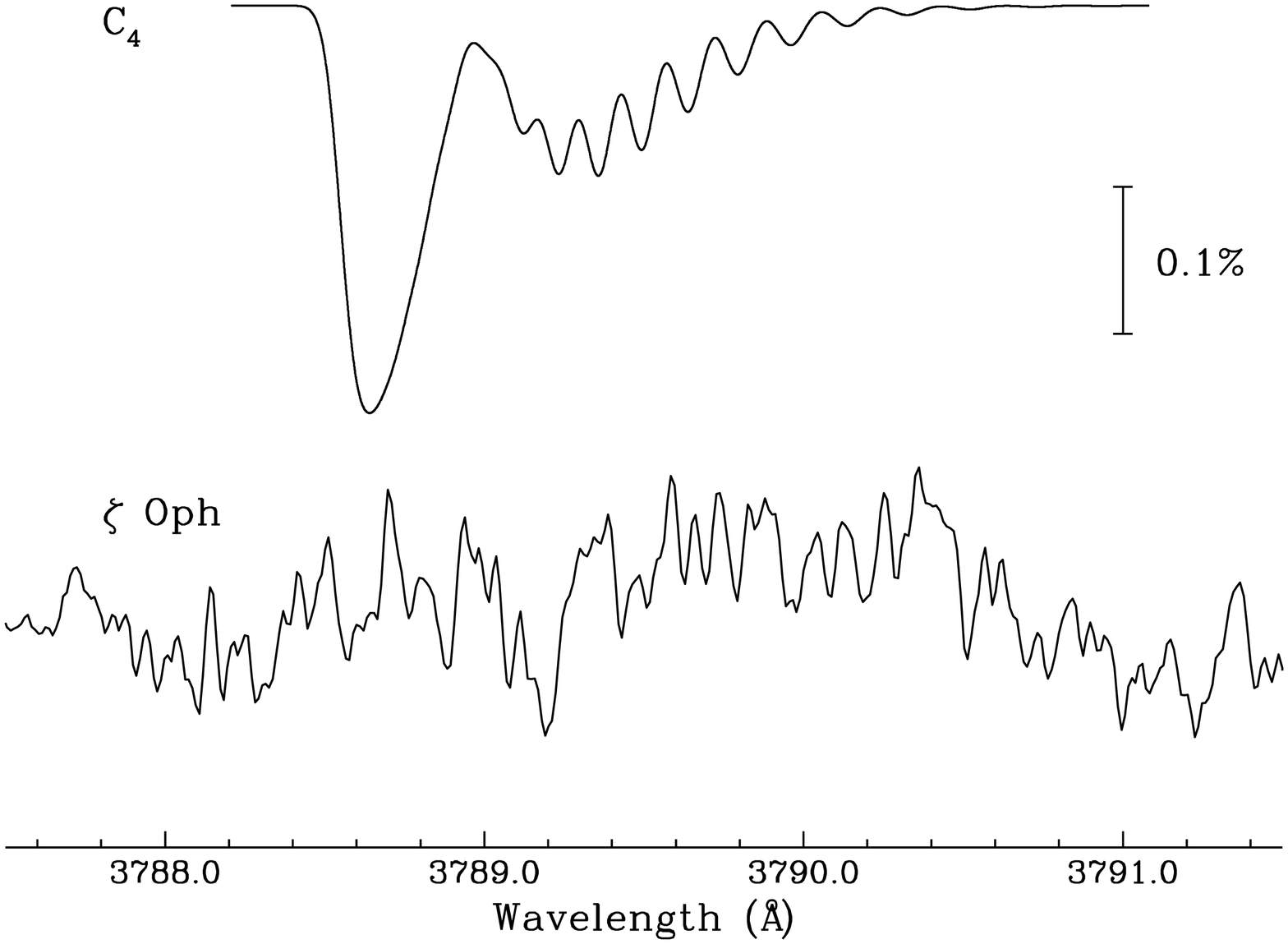,height=13cm,angle=-90}}
\caption{
Comparison of a laboratory spectrum (top) of C$_{4}$ at 3788 \AA\/ from \cite{Lin00} smoothed to a
spectral resolution of 110,000 and compared to the observed spectrum
(lower) of $\zeta$ Oph from Figure 1.
\label{figure2} }
\end{figure}

\begin{figure}
\centerline{\psfig{figure=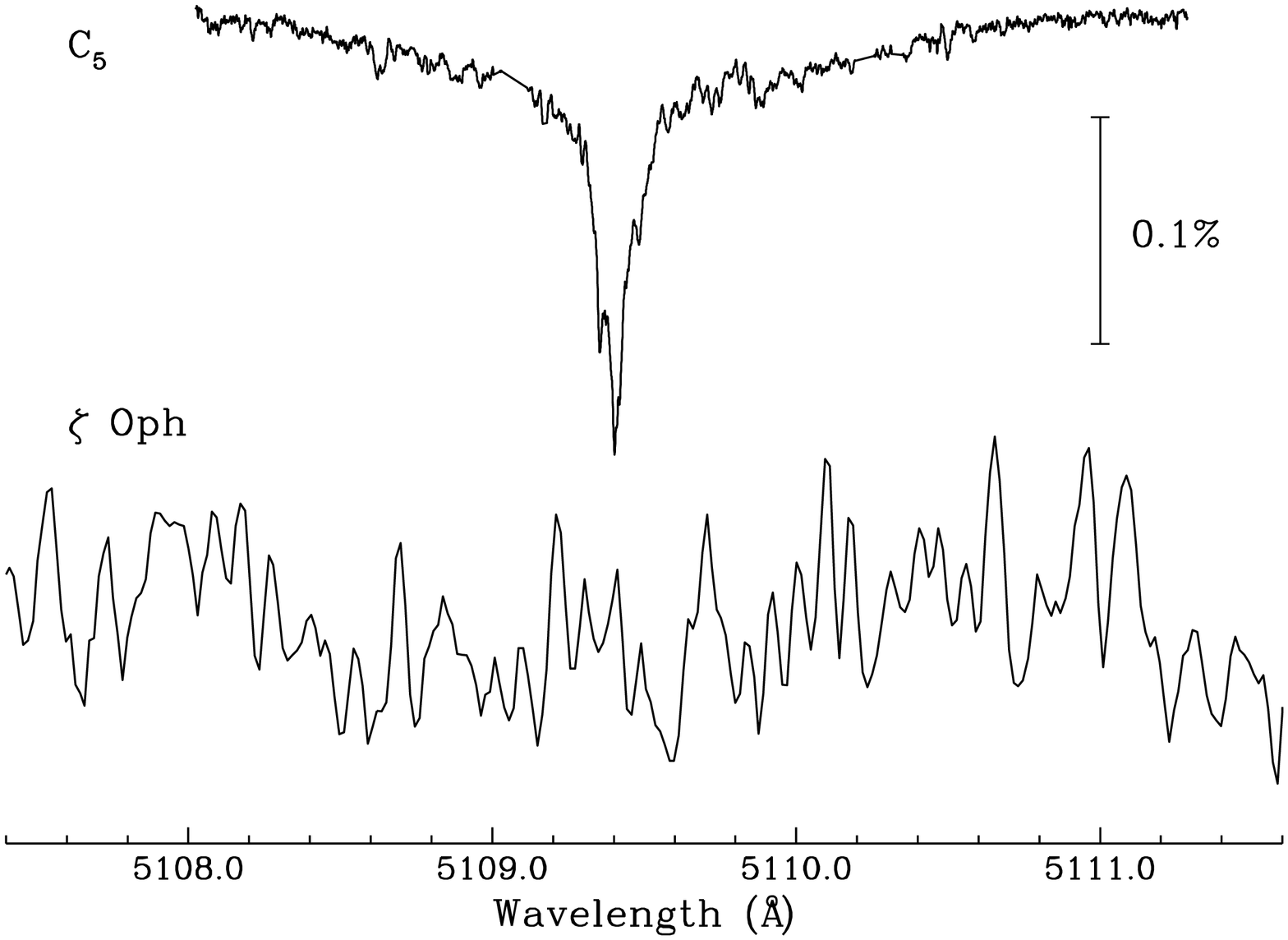,height=13cm,angle=-90}}
\caption{
Comparison of a laboratory spectrum (top) of C$_{5}$ at 5109 \AA\/ from \cite{Mot99} smoothed to a
spectral resolution of 110,000 and compared to the observed spectrum
(lower) of $\zeta$ Oph.
\label{figure3} }
\end{figure}

\end{document}